\documentclass[12pt]{iopart}
\usepackage{graphicx,color}
\expandafter\let\csname equation*\endcsname\relax
\expandafter\let\csname endequation*\endcsname\relax
\usepackage{amsmath}
\usepackage{iopams}
\usepackage{textcomp}

\begin{document}

\title[]{Effects of confinement on thermal stability and folding kinetics in a simple Ising--like model}
\author{M Caraglio$^{1,2,\dagger}$ and A Pelizzola$^{1,2,3,\ddagger}$}
\address{$^1$ Dipartimento di Fisica, CNISM and Center for Computational Studies, Politecnico di Torino, 
Corso Duca degli Abruzzi 24, I-10129 Torino, Italy}
\address{$^2$ INFN, Sezione di Torino, via Pietro Giuria 1, I-10125 Torino, Italy}
\address{$^3$ HuGeF Torino, Via Nizza 52, I-10126 Torino, Italy}
\ead{$^{\dagger}$michele.caraglio@polito.it}
\ead{$^{\ddagger}$alessandro.pelizzola@polito.it}

\begin{abstract}
In cellular environment, confinement and macromulecular crowding play an
important role on thermal stability and folding kinetics of a protein. We
have resorted to a generalized version of the Wako--Sait\^{o}--Mu\~{n}oz--Eaton
model for protein folding to study the behavior of six different protein
structures confined between two walls. Changing the distance $2R$ between the walls,
we found, in accordance with previous studies, two confinement regimes:
starting from large $R$ and decreasing $R$, confinement first enhances the stability
of the folded state as long as this is compact and until a given value of $R$; 
then a further decrease of $R$ leads to a decrease of folding temperature and folding rate.
We found that in the low confinement regime both unfolding temperatures and logarithm
of folding rates scale as $R^{-\gamma}$ where $\gamma$ values lie in between $1.42$ and $2.35$.
\end{abstract}
\noindent{\it protein folding; spatial confinement; Wako--Sait\^{o}--Mu\~{n}oz--Eaton model  \/}

\maketitle

\section{Introduction}

In the past the majority of experiments on protein folding have been
carried out in diluted solutions but in the last two decades it has become
clear that these experiments do not take into account two issues which
arise \textit{in vivo} and whose relevance on thermal stability and
equilibrium rates is not negligible. Namely, crowding and confinement 
\cite{Ellis2001,Minton1997,Minton2000,Ellis2001_COSB}. 
Crowding refers to the fact that about 30\% of cells internal volume is 
occupied by macromolecules such as lipids, carbohydrates and proteins 
themselves \cite{Ellis2001}. This fraction could even reach 
40\% in \textit{E. Coli} \cite{Zimmerman1991}. Confinement is merely a limitation 
in the volume available to the polypeptide chain as naturally occurs in 
the exit tunnel of ribosomes or in the chaperonin cavity.

Studying protein folding properties in a crowded environment is experimentally
possible simply by adding high concentrations of macromolecules to solutions, but 
this approach has problems because of specific interactions which arise
between proteins and crowding agents and because crowding promotes protein--protein
aggregation \cite{Ellis2001}. Based on the idea that the main effect of crowding is 
the reduction of volume available to the protein due to steric constraints, theoretical 
studies and simulations have shown that crowding may be quantitatively mapped onto
confinement as long as crowding agents are modelled as hard spheres and the volume fraction occupied by them does not exceed $10$\% \cite{ThirumalaiPNAS102_2005}. 
Thanks to this mapping, experimental and theoretical studies on confinement may give many hints also for crowding effects.
However the above conditions often does not hold in the cell interior because of too high concentration of agents or presence of macromolecules--protein attractive interaction. In addition, gradients in macromolecule concentrations may exist \cite{Gierasch2011} and, from a more general point of view, crowding is dynamic in nature whereas confinement is static. Thus, the mapping is not close enough to draw a completely satisfactory analogy between crowding and confinement.

An experimental procedure to mimic the effects of confinement,
is the encapsulation of proteins within pores of silica gels 
\cite{Eggers2001,Campanini_2005} or glasses \cite{Ravindra_2004} or polyacrylamide 
gels \cite{Temussi_2004}.
These experiments reported, for most of the considered proteins, an increase in thermal 
stability when they are confined into nanopores. Melting temperature 
($T_{\mbox{\footnotesize{f}}}$) shift is even dramatic in the cases of $\alpha$--lactalbumin 
and RNase A, being as large as about $30$ K \cite{Eggers2001,Ravindra_2004}. On the contrary, recent experiments suggested that crowding influence on stability is modest \cite{Gierasch2011,Gierasch2010}.

The commonly accepted reason for the increase in stability is the change in conformational 
entropy induced by confinement 
\cite{ZhouHX2004,ZhouHX2008,ThirumalaiPNAS100_2003,ThirumalaiPNAS99,Takagi2003,Best2008}. 
Encapsulating the protein in a given volume disallows the most expanded configurations 
of the denatured state ensemble and so indirectly favours more compact structures and, 
among them, the folded state.
The same argument explains also why confinement should lead to an increase in folding rates 
$k_{\mbox{\footnotesize{f}}}$ as long as the nanopore size is large enough to contain 
the folded state and to permit chain reconfigurations around it 
\cite{ZhouHX2004,ZhouHX2008,ThirumalaiPNAS100_2003,ThirumalaiPNAS99,Takagi2003,Best2008,Hartl2006}.

From polymer physics we know that a polymer confined between two (sufficiently close) inert hard walls, behaves like a pancake with the radius of gyration (parallel to the walls) that scales as a power of the number of monomers \cite{DeGennes,Sakaue2006,Molisana1997}. Furthermore, when it is confined within a cage with repulsive 
walls, its free energy follows a simple power law dependence on the size of the cage $R$ 
\cite{DeGennes,Sakaue2006}. Then, as shown by Takagi \textit{et al.} \cite{Takagi2003}, 
folding temperatures and rates should follow the scaling laws 
$\Delta T_{\mbox{\footnotesize{f}}} \sim \Delta \ln k_{\mbox{\footnotesize{f}}} \sim R^{-\gamma}$.

Literature reports many values for the exponent $\gamma$:
for an ideal gaussian chain confined between two walls ($d_{\mbox{\footnotesize{c}}} = 1$), 
in a cylinder ($d_{\mbox{\footnotesize{c}}} = 2$) or in a spherical cavity 
($d_{\mbox{\footnotesize{c}}} = 3$), $\gamma =2$ while for an excluded volume chain 
$\gamma = 5/3$ for $d_{\mbox{\footnotesize{c}}} = 1,2$ and $\gamma = 15/4$ 
for $d_{\mbox{\footnotesize{c}}} = 3$. Using a G\={o}--model $\alpha$--carbon 
representation of proteins and Langevin simulations in a cylindrical cage, Takagi \textit{et al.} 
\cite{Takagi2003} found $\gamma = 3.25 \, \pm \, 0.09$. Best and Mittal \cite{Best2008} 
simulated confinement of protein G and a $3$--helix bundle in different geometries 
and reported that for $d_{\mbox{\footnotesize{c}}} = 1,2$ both values 
$\gamma = 2$ and  $\gamma = 5/3$ are a good estimate of the behavior 
of the two proteins, but they also remarked the fact that it is hard to distinguish 
which value fits best the simulations because least square fitting of power 
laws can produce biased estimates of parameters for small samples. For spherical 
confinement the same authors reported a behavior which is stronger than $\gamma=2$ 
but much weaker than expected behavior for the excluded volume chain ($\gamma = 15/4$).

In the present work we confine a simple Ising--like 
model (WSME model) originally proposed by Wako and Sait\^{o} in 1978 and later 
reconsidered by Mu\~{n}oz and Eaton 
\cite{JPhysSocJpn441931,JPhysSocJpn441939,Nature390,ProcNatlAcad95,ProcNatlAcad96}.
Equilibrium thermodynamics of the model can be solved exactly \cite{PRL88}.
The cluster variation method is exact for this model \cite{JStat2005} and it
successfully describes the kinetics of protein folding \cite{PRL97,JStat2006,JCHEMPHYS126,PRL99}.
More recently it has been proposed a generalized version of the model that permits to reproduce
the general features of mechanical unfolding \cite{PRL98,JCHEMPHYS127} and, through
Monte Carlo simulations, to obtain for some already widely studied proteins
and RNA fragments, unfolding pathways which are consistent with results of experiments
and/or of simulations made with more detailed models \cite{PRL100,PRL103,JCHEMPHYS133,PRE_inpress}.
The model has also been used with success to study folding equilibrium and kinetics and
to mimic mutations of a small ankyrin repeat protein \cite{JCHEMPHYS134}.

We use the confined WSME model to study thermodynamics and kinetics of three ideal
structures and three simple proteins in confining conditions.
The ideal structures are a $10$ residues ideal $\alpha$--helix, a $2$--stranded and
a $3$--stranded ideal $\beta$--sheets each with $7$ residues per strand.
Real structures are a $3$--helix bundle, protein G and its C--terminal $\beta$--hairpin.
The paper is organized as follows: in Sec. 2 we describe the WSME model and its
confinement. Sec. 3.1 focuses on the confinement--induced changes of thermodynamic
stability for the different proteins while Sec. 3.2 deals with kinetics and the expected
increase in folding rates. Some conclusions are drawn in Sec. 4.

\section{The model}

WSME model is a G\={o}--like model in which a given $N$ residues protein
is described by a sequence of $N$ binary variables $m_k$, whose value is $1$
if $k$--th residue is in the native configuration and $0$ otherwise. Two residues
interact only if they and all residues between them are native and only if they are
in contact in the native structure, i.e. they have at least a pair of atoms which
are closer than the threshold length of $4$ \AA \ in the native structure. 
If residues $i$ and $j$ are in contact in the native structure we associate to
them a negative energy $-\varepsilon_{ij}$ (defined as in \cite{ProcNatlAcad96})
and a contact matrix element $\Delta_{ij}=1$. If the two residues are not in
contact $\Delta_{ij}=0$. When the molecule is pulled at its ends by a constant
force $f$, the Hamiltonian reads:
\begin{equation} \label{eq_Ham2}
H \left( \lbrace m_k \rbrace , \lbrace \sigma_{ij} \rbrace \right) = - \sum_{i=1}^{N-1} \sum_{j=i+1}^N \varepsilon_{ij} \Delta_{ij} \prod_{k=i}^j m_k - f L
\end{equation}
where $L=L \left( \lbrace m_k \rbrace , \lbrace \sigma_{ij} \rbrace \right)$ is the
end--to--end length of the protein and $ \lbrace \sigma_{ij} \rbrace $ is a set of
new binary variables that will soon be defined and in which the greater
entropy of non--native states is encoded.

Here and in the following we define a native stretch from residue $i$
to residue $j$ as a sequence of native residues delimited by the two non--native
residues $i$ and $j$. The end--to--end length $L$ is the sum of the native stretches
lengths $l_{ij}$ multiplied by a sign $+1$ or $-1$ (the binary variable $\sigma_{ij}$)
if the stretch is parallel or antiparallel to the direction of the force.
The binary variable $\sigma_{ij}$ thus represents the direction of the stretch from
$i$--th to $j$--th residue.
Using the quantity $S_{ij}=(1-m_i) m_{i+1} \ldots m_{j-1}(1-m_j)$, which is equal to
$1$ if the sequence of residues from $i$ to $j$ is a native stretch and is $0$ otherwise,
and setting the boundary conditions $m_0=m_{N+1}=0$, the length $L$ is defined as:
\begin{equation} \label{eq_L}
L \left( \lbrace m_k \rbrace , \lbrace \sigma_{ij} \rbrace \right) =  \sum_{i=0}^{N} \sum_{j=i+1}^{N+1} \sigma_{ij} l_{ij} S_{ij}
\end{equation}

\begin{figure}[h]
\center
\includegraphics[width=8cm]{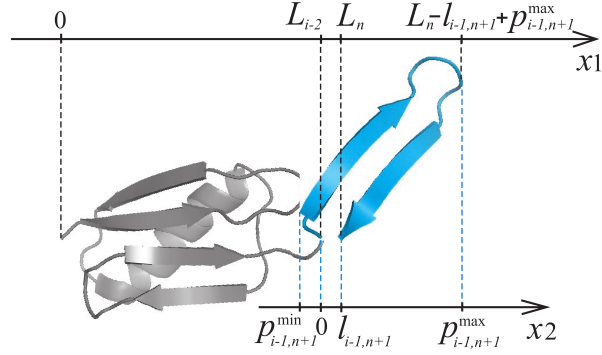}
\caption{Sketch of a configuration with residue $m_{i-1}=0$. Axis $x1$ shows relevant lengths of entire molecule. Axis $x2$ shows relevant lengths of native stretch from ($i-1$)--th to ($n+1$)--th residues.}
\label{fig_lungh}
\end{figure}

The set of all possible lengths $ \lbrace l_{ij} \rbrace $ is obtained directly
from the three dimensional structure deposited in the Protein Data Bank (pdb) as
the distances between the various pairs of central carbon atoms
$ \lbrace \mbox{C}_{\alpha_i} , \mbox{C}_{\alpha_j} \rbrace $. Besides $l_{ij}$,
other two lengths associated to the stretch from the $i$--th to the $j$--th residue
are important for what follows. These are the maximum $p_{ij}^{\mbox{\scriptsize{max}}}$
and the minimum $p_{ij}^{\mbox{\scriptsize{min}}}$ among the distances between
$\mbox{C}_{\alpha_i}$ and the projections of each
$\mbox{C}_{\alpha_k}$ ($i \leq k \leq j$) on the straight line from
$\mbox{C}_{\alpha_i}$ to $\mbox{C}_{\alpha_j}$. Note that, as shown in figure~\ref{fig_lungh}
(axis $x2$), $p_{ij}^{\mbox{\scriptsize{max}}} \geq l_{ij}$ and
$p_{ij}^{\mbox{\scriptsize{min}}} \leq 0 $.

The constrained zero--force partition function
\begin{equation}
Z(L;f=0) = \sum_{ \lbrace m_k \rbrace} \sum_{ \lbrace \sigma_{ij} \rbrace}  \delta \left( L-L(\lbrace m_k \rbrace , \lbrace \sigma_{ij} \rbrace) \right) e^{-\beta H \left( \lbrace m_k \rbrace , \lbrace \sigma_{ij} \rbrace ; f=0 \right)  }
\end{equation}
can be recursively calculated building up the protein residue by residue and
evaluating at each step the partition function $z_n(L)$, where $n$ is the number
of residues achieved at that step (see appendix of~\cite{JCHEMPHYS127} for
detailed calculations).
\begin{equation}\label{eq_RecursiveScheme}
\left\{
\begin{array}{l}
a_n^i (L) = e^{ \beta \chi_{in}} \left[ z_{i-2}(L-l_{i-1,n+1}) + z_{i-2}(L+l_{i-1,n+1}) \right] \\ \quad \\
 z_n(L) = \sum_{i=1}^{n+1} a_n^i (L)
\end{array}
\right.
\end{equation}
Where $\chi_{in}=\sum_{k=i}^{n-1}\sum_{r=k+1}^{n} \varepsilon_{kr}\Delta_{kr}$ is
minus the energy of the native stretch from ($i-1$)--th to ($n+1$)--th residue and the
initial conditions are $z_{-1}(L)=1$ for $L=0$ and $z_{-1}(L)=0$ for $L \ne 0$. The
goal of the recursive scheme is the constrained partition function $Z(L;f=0)$ which
corresponds to $z_N(L)$.
The absolute value of the possible end--to--end lengths of a protein cannot be greater
than $L_{\mbox{\footnotesize{max}}} = \sum_{i=0}^N l_{i,i+1}$, which corresponds to the
length of the molecule in the completely unfolded, fully extended configuration. Thus, 
because of finite resolution of amino acids coordinates in the pdb file
(which is $10^{-3}$ \AA ), $L$ belongs to a finite set of values in the range
$\left[ -L_{\mbox{\footnotesize{max}}}, L_{\mbox{\footnotesize{max}}} \right] $.

\subsection{Confinement of WSME model}

Consider again the recursive scheme of (\ref{eq_RecursiveScheme}) and set the 
starting point of the molecule in the middle of the cage.
In order to confine the protein into a cage of size $2R$ with perfectly
repulsive walls, when adding a native stretch from ($i-1$)--th to ($n+1$)--th
residues (which are respectively at the distances $L_{i-2}$ and $L_n$ from the
N--terminus), one has to require that every residue of this stretch lie inside
the cage. This issue may be solved by considering also the lengths
$p_{i-1,n+1}^{\mbox{\scriptsize{max}}}$ and
$p_{i-1,n+1}^{\mbox{\scriptsize{min}}}$ of the native stretch (see axis $x1$
of figure~\ref{fig_lungh}) and inserting appropriate step functions in the recursive
scheme:
\begin{equation}\label{eq_RecSchemeConf1}
\left\{
\begin{array}{l}
 \begin{array}{lcl}
 a_n^i (L) & = & e^{ \beta \chi_{in}}  \; \times \\
 \end{array} \\
\quad \times \; \left[ z_{i-2}(L-l_{i-1,n+1}) \theta(R-L + l_{i-1,n+1} -p_{i-1,n+1}^{\mbox{\scriptsize{max}}}) \theta(R-l_{i-1,n+1}+L + p_{i-1,n+1}^{\mbox{\scriptsize{min}}}) + \right. \\
\quad \left. + z_{i-2}(L+l_{i-1,n+1}) \theta(R+L + l_{i-1,n+1} - p_{i-1,n+1}^{\mbox{\scriptsize{max}}}) \theta(R-l_{i-1,n+1}-L + p_{i-1,n+1}^{\mbox{\scriptsize{min}}}) \right] \\
\quad \\
 z_n(L) = \sum_{i=1}^{n+1} a_n^i (L)
\end{array}
\right.
\end{equation}
where $\theta$ is the Heaviside step function:
\begin{equation}
\theta(x)=
\left\{
\begin{array}{ll}
1 & \mbox{ if  } x \geq 0 \\
0 & \mbox{ else  }
\end{array}
\right.
\nonumber
\end{equation}

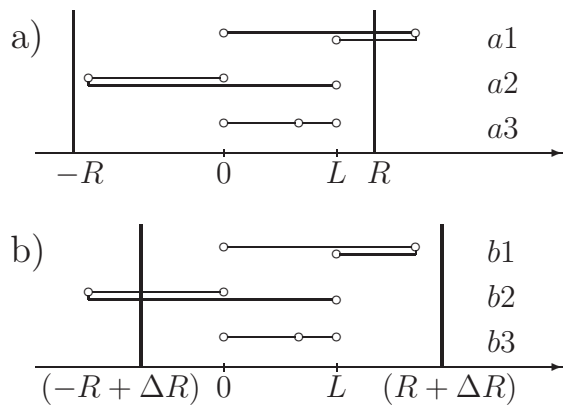
\begin{figure} [htbp] 
\setlength{\unitlength}{1.cm}
\begin{center}

\begin{picture}(7.5,5.5)(0,0)
\put(0,0){\makebox(7.5,5.5){}}

\put(0,4.65){\begin{large} a) \end{large}}
\put(0.5,3.25){\vector(1,0){7.}}
\put(3,3.2){\line(0,1){0.1}}
\put(0.76,2.85){$-R$}
\put(4.9,2.85){$R$}
\put(2.9,2.85){$0$}
\put(4.5,3.2){\line(0,1){0.1}}
\put(4.36,2.85){$L$}

\put(3,4.85){\circle{0.1}}
\put(3.05,4.85){\line(1,0){2.45}}
\put(5.55,4.85){\circle{0.1}}
\put(5.55,4.8){\line(0,-1){0.05}}
\put(5.54,4.75){\line(-1,0){1.}}
\put(4.5,4.75){\circle{0.1}}
\put(6.5,4.65){$a1$}

\put(3,4.25){\circle{0.1}}
\put(2.95,4.25){\line(-1,0){1.7}}
\put(1.2,4.25){\circle{0.1}}
\put(1.2,4.2){\line(0,-1){0.05}}
\put(1.2,4.15){\line(1,0){3.25}}
\put(4.5,4.15){\circle{0.1}}
\put(6.5,4.05){$a2$}

\put(3,3.65){\circle{0.1}}
\put(3.05,3.65){\line(1,0){0.9}}
\put(4,3.65){\circle{0.1}}
\put(4.05,3.65){\line(1,0){0.4}}
\put(4.5,3.65){\circle{0.1}}
\put(6.5,3.45){$a3$}

\put(0,1.8){\begin{large} b) \end{large}}
\put(0.5,0.4){\vector(1,0){7.}}
\put(3,0.35){\line(0,1){0.1}}
\put(0.56,0){$(-R+\Delta R)$}
\put(5.1,0){$(R+\Delta R)$}
\put(2.9,0){$0$}
\put(4.5,0.35){\line(0,1){0.1}}
\put(4.36,0){$L$}

\put(3,2){\circle{0.1}}
\put(3.05,2){\line(1,0){2.45}}
\put(5.55,2){\circle{0.1}}
\put(5.55,1.95){\line(0,-1){0.05}}
\put(5.54,1.9){\line(-1,0){1.}}
\put(4.5,1.9){\circle{0.1}}
\put(6.5,1.8){$b1$}

\put(3,1.4){\circle{0.1}}
\put(2.95,1.4){\line(-1,0){1.7}}
\put(1.2,1.4){\circle{0.1}}
\put(1.2,1.35){\line(0,-1){0.05}}
\put(1.2,1.3){\line(1,0){3.25}}
\put(4.5,1.3){\circle{0.1}}
\put(6.5,1.2){$b2$}

\put(3,0.8){\circle{0.1}}
\put(3.05,0.8){\line(1,0){0.9}}
\put(4,0.8){\circle{0.1}}
\put(4.05,0.8){\line(1,0){0.4}}
\put(4.5,0.8){\circle{0.1}}
\put(6.5,0.6){$b3$}

\thicklines
\put(1,3.25){\line(0,1){1.9}}
\put(5,3.25){\line(0,1){1.9}}
\put(1.9,0.4){\line(0,1){1.9}}
\put(5.9,0.4){\line(0,1){1.9}}

\end{picture}
\end{center}
\caption{\label{fig_mvCage} Three different configurations which would give a contribution to the partition function constrained at length $L$ without any cage.  With cage \textit{a} only configurations $2$ and $3$ contribute. In \textit{b} only configurations $1$ and $3$ contribute.}
\end{figure}

Translational freedom must also be taken into account. To this end, for a given configuration,
instead of considering simply the end-to-end length, it would be better to consider as
the relevant length the distance between the two farthest residues of that configuration.
We call it the configuration effective length. Fixing in the center of the cage the
N--terminus excludes from the partition functions $z_n(L)$ the contribution of some of the
configurations which have an effective length shorter than $2R$ (for example in fig.~\ref{fig_mvCage}a configuration $a1$ has an effective length shorter than configuration $a2$ but the former
is forbidden while the latter is allowed).
Thus, to take into account all the configurations with an effective length
shorter than the cage size, the partition function has to be computed for
different positions of the cage relative to the N--terminus.
The final partition function will be the sum of various partition functions at
different cage positions. Note that some configurations will appear many times
in such a scheme (for example state $a3$ of fig.~\ref{fig_mvCage}a) as a consequence
of their greater translational freedom.

To obtain the final partition function one has to repeat this procedure considering all
the possible positions of the cage relative to the N--terminus, i.e. to start with the
range $\left[ -2R,0 \right] $ and to move the cage with a step $\Delta R$ equal to the
resolution of the $ \lbrace l_{ij} \rbrace$ until the final range $\left[ 0,2R \right] $
is reached. To speed up computations we rounded the lengths to a resolution of $10^{-1}$
\AA~ . For the $3$--helix bundle we checked  that this assumption does not modify the
results through a comparison with results obtained at the resolution of $10^{-3}$ \AA .

\section{Results and discussion}

\subsection{Equilibrium}

\begin{figure} [htbp] 
\setlength{\unitlength}{1.cm}
\begin{center}

\begin{picture}(6,2.2)(0,0)
\put(0,0){\makebox(6,2.2){}}

\multiput(0, 0)(1, 0){7}{\circle*{0.18}}
\multiput(0, 1)(1, 0){7}{\circle*{0.18}}
\multiput(0, 2)(1, 0){7}{\circle*{0.18}}

\multiput(0,0.15)(1,0){6}{\line(0,1){0.15}}
\multiput(0,0.425)(1,0){6}{\line(0,1){0.15}}
\multiput(0,0.7)(1,0){6}{\line(0,1){0.15}}
\multiput(1,1.15)(1,0){6}{\line(0,1){0.15}}
\multiput(1,1.425)(1,0){6}{\line(0,1){0.15}}
\multiput(1,1.7)(1,0){6}{\line(0,1){0.15}}

\thicklines
\put(0,0){\line(1,0){6}}
\put(0,1){\line(1,0){6}}
\put(0,2){\line(1,0){6}}
\put(6,0){\line(0,1){1}}
\put(0,1){\line(0,1){1}}

\end{picture}
\end{center}
\caption{\label{fig_sheet} Ideal antiparallel $\beta$--sheet with $3$ strands. Distance between two consecutive residues is $3.8$ \AA . Dashed lines represent active contacts. For them $\varepsilon_{ij}=1$, while $\varepsilon_{ij}=0$ in other cases.}
\end{figure}
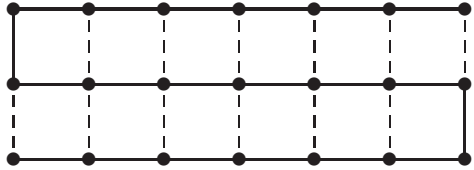

In this study we considered six different structures. Three are real structures:
a $3$--helix bundle (pdb code 1PRB), protein G (pdb code 2GB1) and its final hairpin.
The other three structures are an ideal $\alpha$--helix of ten residues (radius $2.3$
\AA , pitch $5.4$ \AA , $\varepsilon_{ij}=1$ if $j=i+4$ and $\varepsilon_{ij}=0$
otherwise), a $2$--stranded and a $3$--stranded antiparallel $\beta$--sheets with $7$
residues in each strand (the $3$--stranded sheet is drawn in figure~\ref{fig_sheet}).
In the following, code `a010' refers to the ideal $\alpha$--helix, `b207' and
`b307' to the two $\beta$--sheets which have respectively $2$ and $3$ strands
and `GB1h' refers to the final hairpin of protein G.

To study the equilibrium response to confinement of the six structures,
we computed, at different cage sizes $R$, thermodynamic quantities as the Helmholtz
free energy, the specific heat and the average fraction of native residues. For each
structure we varied the distance $2R$ between the walls, in a range from about the
mininum effective length of the completely unfolded state to twice the maximum length
of the completely unfolded state, i.e. from $4$ \AA~ (the distance between two subsequent
amino acids is about $3.8$ \AA ) to $2L_{\mbox{\footnotesize{max}}}$.

If we denote with $L_{\mbox{\scriptsize{N }\footnotesize{eff.}}}$
the effective length of the native state (values are reported in table~\ref{tab_lengths}),
we may naively distinguish between two different confinement
regimes: (\textit{i}) one, for $2R > L_{\mbox{\scriptsize{N }\footnotesize{eff.}}}$,
which disallows the more expanded conformations of the non--native basin but not the
folded state, and (\textit{ii}) the strong confinement regime, for
$2R < L_{\mbox{\scriptsize{N }\footnotesize{eff.}}}$, which forbids also the fully
native state. 

Table~\ref{tab_lengths} also shows the effective length of the unfolded state $L_{\mbox{\scriptsize{U }\footnotesize{eff.}}}$. This is obtained through a Monte Carlo simulation at the unfolding temperature as the average effective length over the configurations belonging to the unfolded basin. Details about Monte Carlo moves will be given in the next section.

\begin{table}[htbp]
\caption{\label{tab_lengths} Native state end--to--end length ($L_{\mbox{\scriptsize{N}}}$), effective length of the native state ($L_{\mbox{\scriptsize{N }\footnotesize{eff.}}}$), maximum length of the fully unfolded state ($L_{\mbox{\footnotesize{max}}}$) and effective length of the unfolded state ($L_{\mbox{\scriptsize{U }\footnotesize{eff.}}}$) for the six different structures.}
\begin{indented}
\item[]
\begin{tabular}{c|cccccc}
\br
 & a010  & b207 & b307 & GB1h & 1PRB & 2GB1 \\ 
\mr
$L_{\mbox{\scriptsize{N}}}$ (\AA )& $14.3$ & $3.8$ & $24.0$ & $6.5$ & $40.0$ & $27.8$ \\ 
$L_{\mbox{\scriptsize{N }\footnotesize{eff.}}}$ (\AA ) &
 $14.3$ & $3.8$ & $24.0$ & $6.6$ & $40.0$ & $29.1$ \\ 
$L_{\mbox{\footnotesize{max}}}$ (\AA )& $34.2$ & $49.4$ & $76.0$ & $63.8$ & $201$ & $212$ \\ 
$L_{\mbox{\scriptsize{U }\footnotesize{eff.}}}$ (\AA ) &
 $14.1$ & $21.8$ & $27.3$ & $24.1$ & $40.5$ & $46.3$ \\ 
\br
\end{tabular}
\end{indented}
\end{table}

\begin{figure}[h]
\center
\includegraphics[width=8cm]{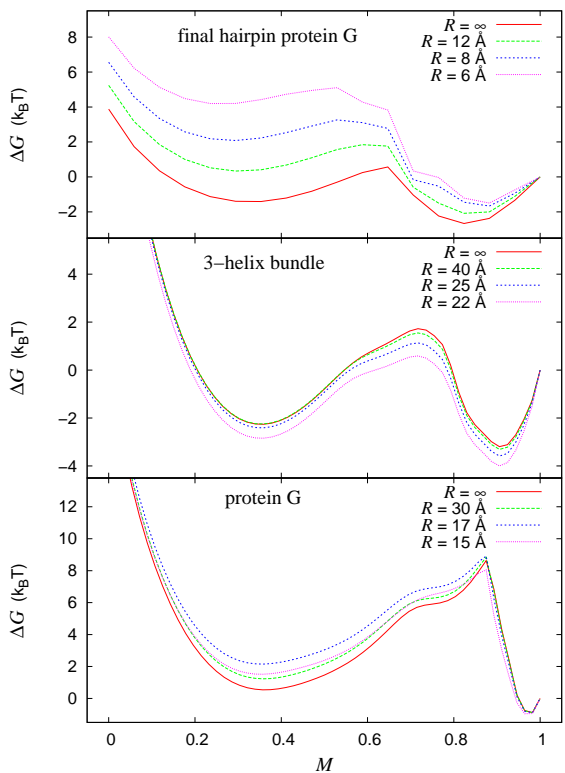}
\caption{Free energy profile in function of the fraction of native residues $M$ at various confinement radius $R$ for the $3$--helix bundel, protein G and its final hairpin. Free energy of completely native state ($M=1$) have been setted to zero.}
\label{fig_freeE}
\end{figure}

Since, without confinement, for a given set of binary variables $ \lbrace m_k \rbrace $,
the model admits $2^{ \sum_{i=1}^N (1-m_i)}$ configurations and this number grows
exponentially with the amount of non--native residues, we may expect that confinement in a cage of size
$R$, with $L_{\mbox{\footnotesize{max}}} > 2R > L_{\mbox{\scriptsize{N }\footnotesize{eff.}}}$,
gives a reduction of conformational entropy which affects more the non--native basin.
Besides, one has to consider translational freedom whose role is to further stabilize the most
compact configurations irrespectively of the fact that they belong or not to the native basin.
Thus a structure with $L_{\mbox{\scriptsize{N }\footnotesize{eff.}}} > L_{\mbox{\scriptsize{U }\footnotesize{eff.}}}$,
as in the case of the ideal $\alpha$--helix, does not undergo any stabilization of the folded
state.
Figure~\ref{fig_freeE} shows the free energy landscapes for the three real structures at 
different confinement sizes (for a better comparison the free energy of the completely folded state
has always been set to zero). For the final hairpin of protein G confinement increases the free energy
of both the native and non--native basin: both native and non--native basin are
destabilized but the latter is more affected. On the contrary, for the $3$--helix bundle
both native and non--native basins are stabilized, with a slightly greater stabilization
with confinement for the native state. Finally, for protein G, only the non--native basin
is destabilized by confinement.

The increased stability of the native state relative to the unfolded state should
result in higher unfolding temperature according to \cite{ThirumalaiPNAS99,Takagi2003}:
\begin{equation}\label{eq_DeltaT}
\frac{T_{\mbox{\footnotesize{f}}} - T_{\mbox{\footnotesize{f}}}^0}{T_{\mbox{\footnotesize{f}}}^0} \propto \left( \frac{2R}{L_{\mbox{\scriptsize{N }\footnotesize{eff.}}}} \right) ^{- \gamma}
\end{equation}
where here, and from now on, we denote with $T_{\mbox{\footnotesize{f}}}^0$ the
unfolding temperature without confinement. For each protein, we have determined
$T_{\mbox{\footnotesize{f}}}$ as the temperature at which the average fraction of
native residues is such that $(M-M_{\infty})/(M_0-M_{\infty}) = 0.5$, where
$M_{\infty} = 1/3$ is the value of $M$ at infinite temperature and $M_0 \approx 1$
is its value at zero temperature.

The ideal $\alpha$--helix is destabilized by confinement already from values of $R$
lower than $R=15$ \AA~ and no enhancement in the unfolding temperature could be detected
for greater values of $R$. Other proteins exhibit an enhancement in their thermal
stability to a different extent depending on their structure: the increase in unfolding
temperatures is of few percents for the $3$--stranded $\beta$--sheet, $3$--helix bundle
and protein G, while for the two $\beta$--hairpins $T_{\mbox{\footnotesize{f}}} \simeq  6.6  \, T_{\mbox{\footnotesize{f}}}^0$ (ideal $2$--stranded $\beta$--sheet)
and $T_{\mbox{\footnotesize{f}}} \simeq  2.7 \, T_{\mbox{\footnotesize{f}}}^0$ (final hairpin of protein G).
Such drastically different
behavior is due to the very short effective lengths of native states of the two hairpins
and to the limitation of the model which projects the positions of all residues on a
single direction and loses information on the real three--dimensional structure. For the
$3$--helix bundle and for protein G, the increases in unfolding temperature correspond
respectively to about $1.5$ K and $9.3$ K. 

Values $R_{\mbox{\footnotesize{I}}}^{\mbox{\footnotesize{eq}}}$ of the cage radius
for which, at equilibrium, unfolding temperature reaches its maximum and the extent
of enhancement are reported in table~\ref{tab_R}.

\begin{table}[htbp]
\caption{\label{tab_R} Values of $R$ for which unfolding temperature reaches its maximum ($T_{\mbox{\footnotesize{f}}}^{\mbox{\scriptsize{max}}}$) and the extent of enhancement. Values of $\gamma$ from fits to (\ref{eq_DeltaT}) and fit ranges. Fits in ranges from $L_{\mbox{\scriptsize{U }\footnotesize{eff.}}} / 2$ to $L_{\mbox{\footnotesize{max}}}$ for `b207' and `GB1h' result in exponents $\gamma \, '$.}
\begin{indented}
\item[]
\begin{tabular}{c|cccccc}
\br
  & b207 & b307 & GB1h & 1PRB & 2GB1 \\ 
\mr
$R_{\mbox{\scriptsize{I}}}^{\mbox{\footnotesize{eq}}}$  (\AA ) &  $2$ & $17$ & $3$ & $25$ & $17$ \\ 
$T_{\mbox{\footnotesize{f}}}^{\mbox{\scriptsize{max}}}$ & 
$6.55 \, T_{\mbox{\footnotesize{f}}}^0$  & 
$1.013 \, T_{\mbox{\footnotesize{f}}}^0$  & 
$2.73 \, T_{\mbox{\footnotesize{f}}}^0$  & 
$1.005 \, T_{\mbox{\footnotesize{f}}}^0$  &
$1.03 \, T_{\mbox{\footnotesize{f}}}^0$   \\ 
\mr
$\gamma$ & $2.14\pm0.03$ & $1.57\pm0.05$ & $2.35\pm0.03$ & $1.50\pm0.05$ & $1.65\pm0.04$ \\
fit range (\AA ) & $[4,50]$ & $[18,76]$ & $[4,64]$ & $[26,201]$ & $[18,212]$ \\
\mr
$\gamma \, '$ & $1.72\pm0.06$ &  & $1.60\pm0.07$ &  &  \\
fit range (\AA ) & $[10.9,50]$ &  & $[12.05,64]$ &  &  \\
\br
\end{tabular}
\end{indented}
\end{table}

\begin{figure}[h]
\center
\includegraphics[width=8cm]{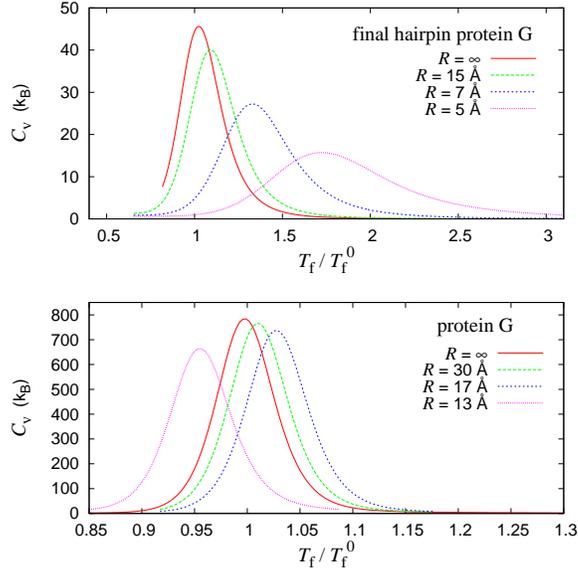}
\caption{Specific heat $C_V = \frac{1}{k_BT^2} \frac{\partial^2 \ln Z}{\partial \beta^2} $ as a function of the temperature at various confinement radius $R$ for protein G and its final hairpin.}
\label{fig_Cv}
\end{figure}

The enhancement in thermal stability can be appreciated in figure \ref{fig_Cv}
where we reported the specific heat as a  function of temperature. The top panel also
shows well another feature of the unfolding phase transition in confined environment
which is a decreased cooperativity with confinement \cite{Takagi2003}.

\begin{figure}[h]
\center
\includegraphics[width=8cm]{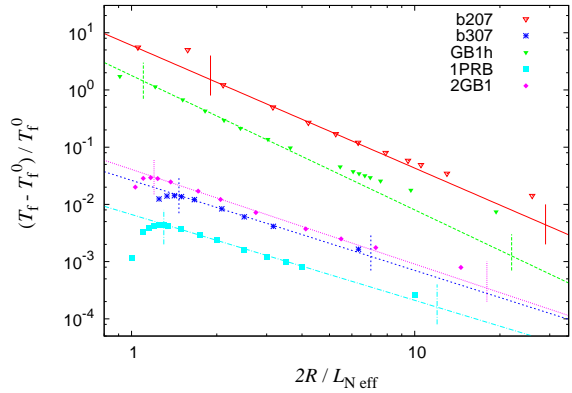}
\caption{Shift in unfolding temperature as a function of confining cage radius $R$. Fits to (\ref{eq_DeltaT}) in ranges reported in table~\ref{tab_R}. The vertical lines represent the ranges spanned by fits.}
\label{fig_fitEq}
\end{figure}

\begin{figure}[h]
\center
\includegraphics[width=8cm]{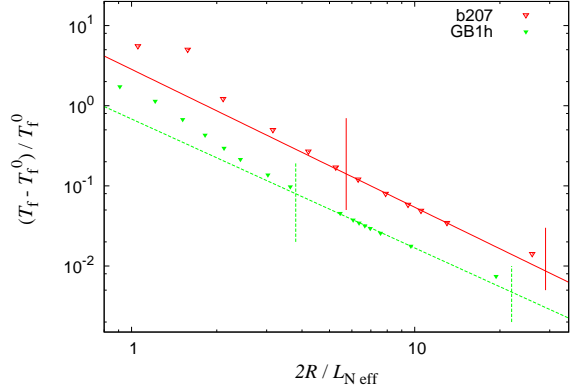}
\caption{Shift in unfolding temperature as a function of confining cage radius $R$ for `b207' and `GB1h'. Fits to (\ref{eq_DeltaT}) in ranges $L_{\mbox{\scriptsize{U }\footnotesize{eff.}}} / 2$ to $L_{\mbox{\footnotesize{max}}}$. The vertical lines represent the ranges spanned by fits.}
\label{fig_fitEq_insert}
\end{figure}

A fit to (\ref{eq_DeltaT}) of unfolding temperatures as a function of $R$
(figure~\ref{fig_fitEq}) yielded exponents $\gamma$ reported in table~\ref{tab_R}.
All values are in between $1.50$ ($3$--helix bundle) and $2.35$ (final hairpin of
protein G). Remarkably, in this range we find also the theoretical values of $\gamma$
for an excluded volume chain confined in a slit or in a cylinder ($\gamma = 5/3$) and
for a gaussian chain in a slit, a cylinder or a sphere ($\gamma = 2$).
Furthermore, a more careful analysis of data in figure~\ref{fig_fitEq} suggested us to fit,
in the case of the $\beta$--hairpins, also in a more limited range of $R$ values going
from $L_{\mbox{\scriptsize{U }\footnotesize{eff.}}} / 2$ to $L_{\mbox{\footnotesize{max}}}$ (figure~\ref{fig_fitEq_insert}). In this very low confinement regime $\gamma = 1.72$ for
the ideal hairpin and $\gamma = 1.6$ for the final hairpin of protein G.


\subsection{Kinetics}

The folding kinetics have been studied by Monte Carlo (MC) simulations in which
a $2$--components ternary variable $(m_k , s_k )$ have been associated to each
residue $k$. If $m_k=1$, $s_k = 0$ while if $m_k=0$, $s_k = \sigma_{kj} = \pm 1$
is the direction of the native stretch from the $k$--th to the $j$-th residue.
A single MC step consists in choosing a residue $k$ with uniform probability among
the $N$ residues and changing $(m_k , s_k )$ variable with equal probability to
any of its other two states. This move is alternated with a $0.1$ \AA~ translation
of the entire protein to the left or to the right with equal probability. Few remarks
are necessary: suppose to have a native stretch from the $i$--th to the $j$-th residue
and to transform the variable $(m_k , s_k ) $, $i<k<j$, from $(1,0)$ to $(0,s_k=\pm 1)$.
The direction of the new native stretch from the $k$--th to the $j$-th residue will be
determined by $s_k$ while the new native stretch from $i$ to $k$ will inherit the
direction of the old one from $i$ to $j$. If instead we move the state of $k$--th
residue from $(0,\pm1)$ to $(1,0)$, two native stretches merge into one with direction
equal to the direction of the first old native stretch. At each MC step confinement
requirements must be checked.

Changes in folding rates have been estimated \cite{Best2008} from Kramers kinetics, $k_{\mbox{\footnotesize{f}}}
\propto D \exp(-\Delta G_{\mbox{\scriptsize{U}}}^{\mbox{\scriptsize{barrier}}}/k_BT)$,
where $\Delta G_{\mbox{\scriptsize{U}}}^{\mbox{\scriptsize{barrier}}}$ is the free
energy difference between the transition state and the unfolded state and $D$ is a
diffusion coefficient. Because the unfolded state is destabilized by confinement, the
free energy barrier dividing the unfolded from the native state is smaller. Assuming
that the free energy of the latter and the diffusion constant are not affected by
confinement and that the free energy of the unfolded state grows by a term
$\sim T(R/L_0)^{-\gamma}$ \cite{DeGennes,Sakaue2006} leads to the scaling law:
\begin{equation}\label{eq_Deltak}
\ln \left( \frac{k_{\mbox{\footnotesize{f}}}}{k_{\mbox{\footnotesize{f}}}^0}   \right)  \propto \left( \frac{2R}{L_{\mbox{\scriptsize{N }\footnotesize{eff.}}}} \right) ^{- \gamma}
\end{equation}

We determined folding rates as the inverse of mean first passage times by using $10^4$ folding 
trajectories. First passage time is defined as the time at which, starting from a
random unfolded configuration, the weighted fraction of native contacts 
($Q = \sum_{i=1}^{N-1}\sum_{j=i+1}^N \varepsilon_{ij}\Delta_{ij}\prod_{k=i}^j m_k
/ \sum_{i=1}^{N-1}\sum_{j=i+1}^N \varepsilon_{ij}\Delta_{ij}$) catches up
with the threshold $0.9$, which ensures the protein has reached the folded state
and has not got stuck in some intermediate. Temperature has been set to
$0.9 \, T_{\mbox{\footnotesize{f}}}^0$.

\begin{table}[htbp]
\caption{\label{tab_k} Values of $R$ for which the folding rate reaches its maximum $k_{\mbox{\footnotesize{f}}}^{\mbox{\scriptsize{max}}}$  at $T=0.9 \, T_{\mbox{\footnotesize{f}}}^0 $ and the extent of enhancement. Values of $\gamma$ from fits to (\ref{eq_Deltak}) in the reported ranges.}
\begin{indented}
\item[]
\begin{tabular}{c|cccccc}
\br
  & b207 & b307 & GB1h & 1PRB & 2GB1 \\ 
\mr
$R_{\mbox{\scriptsize{I}}}^{\mbox{\footnotesize{kin}}}$  (\AA ) &  $19$ & $23$ & $12$ & $19$ & $18$ \\ 
$k_{\mbox{\footnotesize{f}}}^{\mbox{\scriptsize{max}}}$ & 
$1.13 \, k_{\mbox{\footnotesize{f}}}^0$  & 
$1.13 \, k_{\mbox{\footnotesize{f}}}^0$  & 
$1.46 \, k_{\mbox{\footnotesize{f}}}^0$  & 
$1.50 \, k_{\mbox{\footnotesize{f}}}^0$  &
$2.35 \, k_{\mbox{\footnotesize{f}}}^0$   \\ 
\mr
$\gamma$ & $1.42\pm0.20$ & $1.53\pm0.33$ & $1.54\pm0.11$ & $1.71\pm0.08$ & $1.67 \pm 0.07$ \\
fit range (\AA ) & $[19,50]$ & $[23,76]$ & $[14,64]$ & $[22,201]$ & $[20,212]$ \\
\br
\end{tabular}
\end{indented}
\end{table}

When decreasing $R$, folding is accelerated until a certain size $R_{\mbox{\scriptsize{I}}}^{\mbox{\footnotesize{kin}}}$ is reached, then folding rates start to decrease. Table~\ref{tab_k} reports $R_{\mbox{\scriptsize{I}}}^{\mbox{\footnotesize{kin}}}$ values and the maximum extent of folding rates enhacement. For the $\beta$--hairpins, the drastic difference between $R_{\mbox{\scriptsize{I}}}^{\mbox{\footnotesize{kin}}}$ and $R_{\mbox{\scriptsize{I}}}^{\mbox{\footnotesize{eq}}}$ is likely due to the fact that for very small confining cages, even if the native state is not compromised, the structure is squeezed so much that chain reconfigurations towards the folded state become difficult. The same reason should explain the small differences between $R_{\mbox{\scriptsize{I}}}^{\mbox{\footnotesize{kin}}}$ and $R_{\mbox{\scriptsize{I}}}^{\mbox{\footnotesize{eq}}}$ of other structures.

\begin{figure}[h]
\center
\includegraphics[width=8cm]{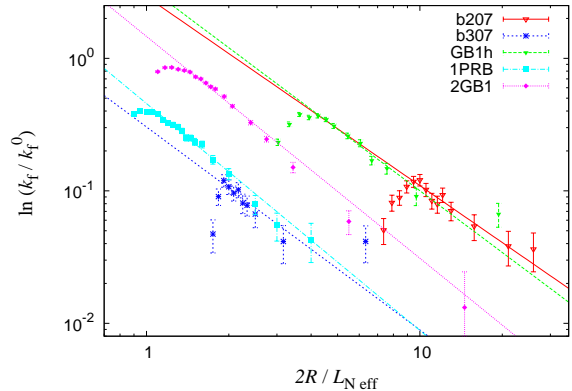}
\caption{Shift in folding rates at $T=0.9 \, T_{\mbox{\footnotesize{f}}}^0 $ as a function of confining cage radius $R$. Fits to (\ref{eq_Deltak}) in ranges reported in table~\ref{tab_k}.}
\label{fig_fitKin}
\end{figure}

Table~\ref{tab_k} also reports the $\gamma$ values obtained through a fit to (\ref{eq_Deltak}) while figure~\ref{fig_fitKin} shows the folding rates behavior together with fit lines. If for the two hairpins we considered the very--low confinement regime, exponents $\gamma$ relative to folding rates are comparable with their equilibrium counterparts.

\section{Conclusions}

We have investigated the effects of confinement on protein thermal stability
and folding kinetics using a simple Ising--like model that we have contributed
to develop and validate in recent years and now properly modified to include
confinement of a polypeptide chain into a slit. To study thermal stability we
have made use of the property of the model to be exactly solvable at equilibrium,
while to study folding rates behavior we have resorted to Monte Carlo simulations.
Notwithstanding the simplicity of the model and its unidimensionality, we
obtained results which follow the general trend of previous experimental studies \cite{Campanini_2005,Ravindra_2004,Temussi_2004,Hartl_2006} and simulations
\cite{ThirumalaiPNAS99,Takagi2003,Best2008}: provided the native state is compact,
when reducing the space available to a given protein, both unfolding temperature
$T_{\mbox{\footnotesize{f}}}$ and folding rate $k_{\mbox{\footnotesize{f}}}$ grow
until a certain confinement size which depends on the protein. If the confinement
size is further decreased, unfolding temperatures and folding rates decrease.
Furthermore, our results also support the theoretical prediction
\cite{ZhouHX2004,Takagi2003,Best2008} that enhancement depends on the confinement
size $R$ by the scaling law $\Delta T_{\mbox{\footnotesize{f}}} \sim \Delta
\ln k_{\mbox{\footnotesize{f}}} \sim R^{-\gamma}$.

Among the six different protein structures studied in this work, one, a $10$--residues
ideal $\alpha$--helix, does not show any enhancement of folding temperature and rate
because its native state cannot be considered compact if compared to the average
unfolded state. For the other five structures we found that exponents $\gamma$ lie in
between the upper and lower values of $2.35$ and $1.42$ and that those obtained for
unfolding temperatures from exact solutions at equilibrium are consistent with those
obtained for folding rates enhancement by Monte Carlo simulations.


Theoretical values of $\gamma$ ($\gamma = 5/3$ for a chain with excluded volume confined into a slit or a cylinder and $\gamma = 2$ for a gaussian chain into a slit, a cylinder or a sphere) are not directly comparable to the results of our model, which differs from these theories both for the geometry (our chain is neither self-avoiding nor gaussian) and for the presence of specific interactions, which are neglected by these theories. Nevertheless, our results, both from thermodynamics and kinetics, for $\gamma$, are in the same range as the theoretical ones.

Furthermore, for a $3$--helix bundle and for protein G, our results are consistent
with those obtained through a more realistic model by Best and Mittal \cite{Best2008}
for confinement of the same proteins into a slit: $\gamma$ values are consistent and
also the maximum enhancement extents of folding temperatures and folding rates are in
good accordance. The two model also agree in the fact that protein G is more affected
by confinement but there is no accordance on the confinement radius at which the
$3$--helix bundle reaches its maximum folding temperature and its maximum folding rate.

\section*{Acknowledgments}
MC thanks Alberto Imparato for allowing the use of cpu resources during his staying in Aarhus university.

\section*{References}

\bibliographystyle{unsrt} 
\bibliography{conf_ReSub}
\end{document}